\documentclass[showpacs,prl,twocolumn,superscriptaddress]{revtex4}
\usepackage{amsmath}
\usepackage{graphicx}
\usepackage{dcolumn}
\usepackage{bm}

\begin{document}
\title{Mesoscopic continuous and discrete channels for quantum information transfer}
\author{Ferdinando  de Pasquale}
\author{Gian Luca Giorgi}
\affiliation{INFM Center for Statistical Mechanics and Complexity}
\affiliation{Dipartimento di Fisica, Universit\`{a} di Roma La
Sapienza, Piazzale A. Moro 2, 00185 Roma, Italy}
\author{Simone Paganelli}
\email{simone.paganelli@roma1.infn.it} \affiliation{Dipartimento di
Fisica, Universit\`{a} di Roma La Sapienza, Piazzale A. Moro 2,
00185 Roma, Italy} \affiliation{Dipartimento di Fisica,
Universit\`{a} di Bologna, Via Irnerio 46, I-40126, Bologna, Italy}

\begin{abstract}
We study the possibility of realizing perfect quantum state transfer
in mesoscopic devices. We discuss the case of the Fano-Anderson
model extended to two impurities. For a channel with an infinite
number of degrees of freedom, we obtain coherent behavior in the
case of strong coupling or in weak coupling off-resonance. For a
finite number of degrees of freedom, coherent behavior is associated
to weak coupling and resonance conditions.

\end{abstract}
\maketitle

Quantum information and quantum communication require the ability of
manipulating and transfer qubits in the space \cite{nielsen}.
Quantum state transfer (QST) can be realized by teleportation
\cite{bennett}, using flying qubits \cite{cirac}, or through quantum
channels. When the information has to be processed in devices
smaller than typical optical wavelengths for flying qubits, quantum
channels are preferable. They can be based on solid state devices or
on confined radiation fields. QST among optical cavities, as
proposed by Cirac {\it et al.} some years ago \cite{cirac}, is
possible due to the fact that each atom inside the cavity interacts
only with a nearly monochromatic photon of the radiation field, and
that photon can be transmitted unchanged to a distant site, before
interacting with another atom in a second cavity. Generally, in
mesoscopic devices, an interaction localized in the space involves
all the modes of the support and the state reconstruction is
affected by interference.

The use of local excitations in quantum chains, first suggested by
Bose \cite {bose}, is indeed far from being optimal, due to quantum
diffusion \cite {osborne,giampaolo}. Different physical realizations
of quantum channels have been suggested: ferromagnetic spin chains
\cite{bose,subry}, Josephson arrays \cite{romito},
nanoelectromechanical oscillators \cite{eisert}. Quantum diffusion
appears in each of these models. Ideally, this drawback can be
overcome by using parallel chains and conditional gates \cite
{burgath} or through the adoption of engineered couplings between
the nodes of the network \cite{datta}. Practically, these solutions
are rather complicated to be realized. Recently, two schemes have
been proposed in which the quantum chain act as a quantum bus, and
both the encoding site and the decoding site are external and
coupled locally with two different points of the chain
\cite{plenio,polacchi}. In Ref.\cite{plenio} the coherent behavior
of the channel is associated with a simple model which takes into
account only the center of mass mode of the chain. Numerical and
analytical considerations show that the transmission in a
finite-size chain can be nearly perfect and the speed of propagation
is independent from the distance between sender and receiver, but
only depends on the total length of the chain.

The latter is one of possible models, where two (or more) quantum
systems interact via an intermediate channel characterized by many
degrees of freedom.

In this Letter we study the general conditions under which such
systems exhibit quantum oscillations, and therefore are suitable for
QST, in the limit of discrete or continuous channel.

For the sake of concreteness, we consider the Fano-Anderson model
\cite {fano,anderson} extended to two impurities:
\begin{eqnarray}
H=\sum_{k}\epsilon _{k}c_{k}^{\dagger }c_{k}+\Omega \left(
c_{A}^{\dagger }c_{A}+c_{B}^{\dagger }c_{B}\right)\nonumber\\
-\frac{g}{\sqrt{N}}\sum_{k}\left[ c_{k}^{\dagger }\left(
c_{A}+e^{ikL}c_{B}\right) +H.c.\right] . \label{fanoanderson}
\end{eqnarray}
We have two quantum systems ($A$ and $B$) with creation and
annihilation operators $c_{A}^{\dagger }$,$c_{A}$,$c_{B}^{\dagger
}$,$c_{B}$, a chain with $N$\ modes, described by $c_{k}^{\dagger }$
$\left( c_{k}\right) $ which creates (annihilates) an excitation in
the mode $k$, and \ interaction with the modes and $A$ and $B$ which
amounts to tunneling processes in the case when both $A$ and $B$ are
associated with a solid state tight binding model, or to a transfer
of energy when $A$ and $B$ are atomic systems interacting with a
radiation field. The coupling constant $g$ measures the strength of
the interaction and the phase factor $\exp \left( ikL\right) $ takes
into account the distance $L$ between $A$ and $B$. In the case of a
continuous spectrum, sums must be thought as integrals. Due to the
quadratic nature of the Hamiltonian, the evolution equation of each
operator is independent from the corresponding quantum statistics.
Then, the model works for fermions as well as for bosons. All the
characteristics of the system are synthesized by the energy
dispersion $\epsilon _{k}$.

In the case of continuum of states, possible candidates as
mesoscopic channels are conductors in the tight binding limit or
one-dimensional wires with magnetic edge states \cite{milburn},
where there are experimental proofs of coherent hopping with quantum
dots \cite{vandervaart,kirczenow}. As far as discrete sets of states
are considered, the model is suitable to be implemented by arrays of
quantum dots, or by nanoelectromechanical oscillators, or by a
radiation confined in a finite-size cavity. An experimental evidence
of coherent oscillations in an all solid state realization of a
Jaynes-Cummings-like scheme has been recently reported
\cite{wallraff}.

We will show that coherent oscillations between $A$ and $B$ can be
achieved using both continuous and discrete channels. In particular,
discrete channels are suitable for our purposes when $A$ and $B$ are
weakly coupled with the chain and $\Omega$ is resonant with one of
its eigenvalues $\epsilon _{k}$. In this situation, only the
resonant modes play a significant role and the effective Hamiltonian
is that of a few-body problem. The same behavior can be attained
with continuous channels in the case of strong coupling, or, in the
weak coupling limit, whenever $\Omega$ lies outside the energy band.

Let us start with an initial state where an excitation is present in
the impurity $A$ and both the second impurity and the channel are in
their respective vacuum states: $\left| \psi _{in}\right\rangle
=c_{A}^{\dagger } \left| 0\right\rangle $. Writing the Heisenberg
equations and replacing the operators with their Laplace transform
$\tilde{c}^{\dagger }\left( \omega \right)
=\int_{0}^{\infty }e^{i\omega t}c^{\dagger }\left( t\right) dt$, assuming $%
\hbar =1$, we find
\begin{widetext}
\begin{equation}
\tilde{c}_{A}^{\dagger }\left( \omega \right) =\frac{i}{D\left(
\omega
\right) }\left\{ \left[ \omega -\Omega -\Lambda _{0}\left( \omega \right) %
\right] \left( c_{A}^{\dagger
}-\frac{g}{\sqrt{N}}\sum_{k}\frac{1}{\omega -\epsilon
_{k}}c_{k}^{\dagger }\right) +\Lambda _{L}\left( \omega \right)
\left( c_{B}^{\dagger
}-\frac{g}{\sqrt{N}}\sum_{k}\frac{e^{ikL}}{\omega -\epsilon
_{k}}c_{k}^{\dagger }\right) \right\} ,  \label{ca}
\end{equation}
\end{widetext}
where
\begin{equation}
\Lambda _{d}\left( \omega \right) =\frac{g^{2}}{N}\sum_{k}\frac{e^{ikd}}{%
\omega -\epsilon _{k}},
\end{equation}
and $D\left( \omega \right) =\left[ \omega -\Omega -\Lambda
_{0}\left(
\omega \right) \right] ^{2}-\Lambda _{L}^{2}\left( \omega \right) $. In Eq. (%
\ref{ca}), terms in $c_{k}^{\dagger }$ , due to the presence of the
excitation in the channel, introduce noise and limit the efficiency
of the channel.

Studying the zeroes of the spectral function $D\left( \omega \right)
$, we extract all information about the system. We can assume
$\epsilon _{k}=-w\cos ka$, as usual when treating solids with many
atoms and lattice constant $a$, and interpret $A$ and $B$ as
impurity states. Here, $2w$ is the bandwidth and $k$ is defined in
the first Brillouin zone limited by $0$ and $2\pi $: $k=2\pi n/N$,
where $n$ is any integer between $0$ and $N-1$.
Without loss of generality, we shall assume throughout the paper $a=1$ and $%
w=1$. It can be shown that, in this case,
\begin{equation}
\Lambda _{d}\left( \omega _{0}\right) =\frac{g^{2}}{\left( \omega
^{2}-1\right) ^{1/2}} \frac{  K_{-d}\left( \omega \right)
+K_{N+d}\left( \omega \right) }{1-K_{N}\left( \omega \right)},
\end{equation}
where $K_{d}\left( \omega \right) =%
\left[ -\omega +\left( \omega ^{2}-1\right) ^{1/2}\right] ^{d}$.

In order to evaluate its zeroes, the spectral function can be
decomposed in two factors: $D(\omega)=D_+(\omega)D_-(\omega)$, where
\begin{eqnarray}\label{laformuletta}
\begin{split}
D_{\pm }\left( \omega \right) &=\omega -\Omega -\frac{g^{2}}{\left(
\omega ^{2}-1\right) ^{1/2}}\\&\times\frac{1+K_{N}\left( \omega
\right)\pm\left[K_{-L}\left( \omega \right)+K_{N+L}\left( \omega
\right)\right]}{1-K_{N}\left( \omega \right)}. \end{split}
\end{eqnarray}
The analytic structure of Eq. (\ref{ca}) consists in $2(N+1)$ real
poles for every finite $N$, and, in the limit $N\rightarrow \infty
$, only $4$ real poles, related to the band extrema, remain, and
poles inside the energy band are substituted by a cut. These exact
results are compared, in the following, with the direct numerical
solution of the evolution equation.

We start from the weak coupling regime ($g\ll 1$). The zeroes of Eq.
(\ref{laformuletta}) can be calculated by iterating the zero order
solution $\omega =\omega _{0}$ obtained in the limit $g\rightarrow
0$.

\begin{figure}[htbp]
\begin{center}
\includegraphics[height=9cm,angle=-90]{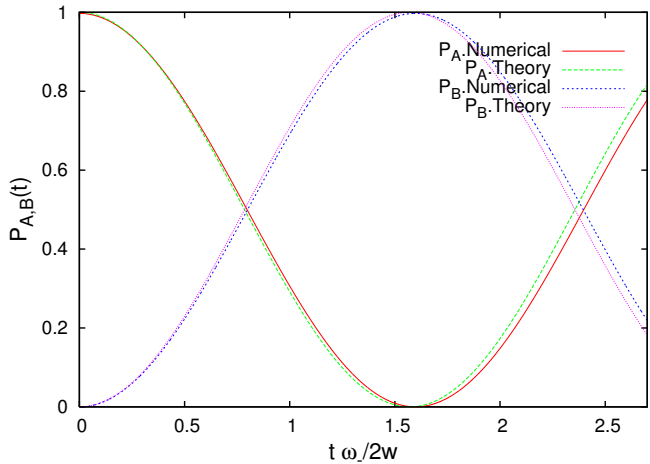}
\end{center}
\caption{Time evolution of the occupation probabilities of $A$ and
$B$ ($P_A$ and $P_B$) in weak coupling and off-resonance. The
coupling strength is $g=0.05$, the impurities' energy is
$\Omega=1.5$, the number of the channel's elements is $N=30$, and
the distance between A and B is $L=6$. Here are reported both the
numerical (exact) and theoretical curves.} \label{fig:fig1}
\end{figure}

First, we assume $\Omega $ outside the energy band: $\left| \Omega
\right| >1$. In this case the zero order solution is $\omega
_{0}=\Omega $ and, by iteration,
\begin{equation}
\omega _{1,2}=\Omega +\Lambda _{0}\left( \Omega \right) \pm \Lambda
_{L}\left( \Omega \right) .
\end{equation}
All roots are real and oscillations are expected. Residues
associated to poles $\omega _{1}$\ and $\omega _{2}$ in
Eq.(\ref{ca}) are obtained neglecting terms in powers of order
$g^{2}$. In such limit we find that all the spectral weight is
concentrated on the impurities' modes. Then, we obtain a coherent
oscillation between the two impurities:
\begin{equation}
c_{A}^{\dagger }\left( t\right) =e^{^{-\frac{i\omega
_{+}t}{2}}}\left(
\cos \frac{\omega _{-}t}{2}c_{A}^{\dagger }-i\sin \frac{%
\omega _{-}t}{2}c_{B}^{\dagger }\right) ,  \label{rabi}
\end{equation}
where $\omega _{+}=2\left[ \Omega +\Lambda _{0}\left( \Omega \right)
\right] $. and $\omega _{-}=2\Lambda _{L}\left( \Omega \right) $.\
In the limit of infinite number of modes, $\omega _{+}=2\Omega
+2g^{2}/\left( \Omega ^{2}-1\right) ^{1/2}$ and $\omega
_{-}=2g^{2}\left[ \Omega -\sqrt{\Omega ^{2}-1}\right]
^{L}/\sqrt{\Omega ^{2}-1}$. These solutions illustrate that the open
system $A+B$ experiences a Rabi oscillation, and actually behaves as
a closed one. Then, the system is suitable for QST or to create
entanglement. In the case discussed above the dependence on the
size-system is not crucial and the continuous limit is achieved even
for not very large values of $N$. In Fig. \ref{fig:fig1} we report
the probabilities of the excitation to be localized either on the
first impurity or on the second one.

The discussion becomes more interesting when $\left| \Omega \right|
<1$. In this  case it can be useful to introduce an auxiliary
complex variable $\gamma$ defined by $\omega=-\cos{\gamma}$, with
the constraint that $0\leq Re\{\gamma\}\leq \pi$. So
\begin{multline}\label{laformula}
D_{\pm}(\gamma)=\cos {\gamma}-\cos{\Gamma}\\
+g^2\frac{1}{\sin{\gamma}\sin{\gamma N/2}} \left[\cos{\gamma N/2}\pm
\cos{\gamma (L+N/2}\right)],
\end{multline}
having defined $\Omega=-\cos{\Gamma}$. At the resonance, $\Omega$
coincides with one of the unperturbed poles.

Since in the weak coupling limit the original energy levels are
slightly modified, it is reasonable to assume that the resonant ones
give the main contribution to the evolution and an expansion around
them can be done. We write $\gamma=\Gamma+\delta$, with $\delta$
expected to vanish in the limit of $g=0$. Then
\begin{equation}\label{expansion}
    \delta\simeq \frac{g^2}{\sin^2{\Gamma}} \left[\cot{\frac{\delta
    N}{2}}\left(1\pm \cos{\Gamma L}\right)\mp \sin{\Gamma L}\right].
\end{equation}

Two different regimes appear for $\delta N \gg 1$ or $\delta N \ll
1$. In the first case the system is well approximated by its
continuum limit, obtained replacing $\cot{\delta N/2}$ with $-i\
sign\{Im\{\delta\}\}$. It is easy to show that Eq. (\ref{expansion})
does not provides polar solutions, but only singularities deriving
from the cut. Under these conditions, the excitation diffuses in the
channel and the QST efficiency is lost.

\begin{figure}[htbp]
\begin{center}
\includegraphics[height=9cm,angle=-90]{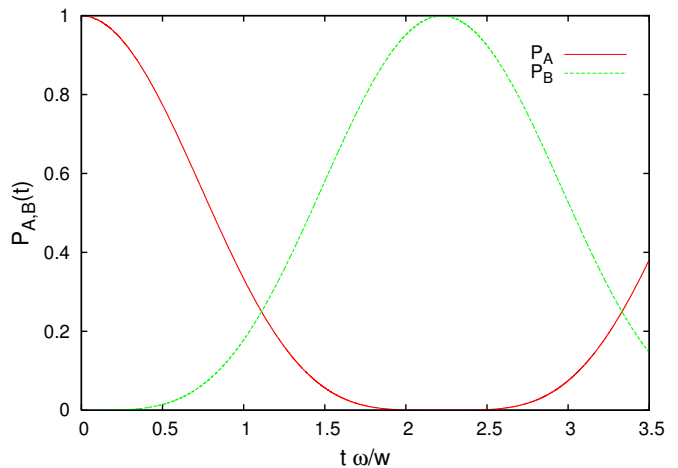}
\end{center}
\caption{Numerical simulation of the evolution of
$P_A\left(t\right)$ and $P_B\left(t\right)$ in weak coupling and
resonance with the following parameters: $g=0.01$, $\Omega=0$,
$N=16$, and $L=8$. The time is normalized with respect to
$\omega=\sqrt{2}g/\sqrt{N}$. The theoretical behavior, calculated in the
text, coincides perfectly with the numerical one.} \label{fig:fig2}
\end{figure}

On the other hand, when $\delta N \ll 1$ the cotangent in
Eq.(\ref{expansion}) is expanded into $2/\left(\delta N\right)$ and
$\sin{\Gamma L}$ is negligible. The solutions are then
$\delta_1^\pm=\pm g \sqrt{2\left(1-\cos{\Gamma L}\right)/N
\sin^2{\Gamma}}$ and $\delta_2^\pm=\pm g \sqrt{2\left(1+\cos{\Gamma
L}\right)/N \sin^2{\Gamma}}$. The time evolution of $c_{A}^{\dagger
}$ looks very simple when $\Omega =0$ and $L$ is even: in such a
case
\begin{eqnarray}
c_{A}^{\dagger }\left( t\right) =\cos
^{2}\frac{gt}{\sqrt{N}}c_{A}^{\dagger
}+\left( -1\right) ^{1+L/2}\sin ^{2}\frac{gt}{\sqrt{N}}c_{B}^{\dagger }\nonumber\\+\frac{i%
}{2}\sin \frac{2gt}{\sqrt{N}}\left( c_{\bar{k}}^{\dagger }+c_{-\bar{k}%
}^{\dagger }\right), \label{simil}
\end{eqnarray}
where $ \pm\bar{k}$ are the modes in resonance with $\Omega=0$. This
formula shows that, despite the non vanishing probability of finding
the excitation in the channel, perfect QST is achieved. In Fig.
\ref{fig:fig2} we report the time evolution of $P_A$ and $P_B$,
which represent the occupation probabilities of $A$ and $B$. On the
other hand, assuming $L$ odd, the second impurity is never
populated:
\begin{equation}
c_{A}^{\dagger }\left( t\right) =\cos gt\sqrt{\frac{2}{N}}
c_{A}^{\dagger }+\frac{i}{\sqrt{2}}\sin gt\sqrt{\frac{2}{N}}\left( c_{\bar{%
k}}^{\dagger }+c_{-\bar{k}}^{\dagger }\right) .
\end{equation}
The result of Eq. (\ref{simil}) is somewhat similar to that obtained
in Ref. \cite{plenio}, showing an efficiency of transfer independent
(limiting ourselves to even values of $L$) from the distance. The
condition $\delta N \ll 1$ (or $g\sqrt{N}\ll 1$) can be interpreted
as follows: the interaction splits the resonant pole in two levels
with an energy separation of the order of $g/\sqrt{N}$, while the
energy spacing between different modes is about $1/N$. If none of
the other modes falls inside this energy interval, then the
excitation interacts effectively only with the resonant modes and
the coherent behavior appears. Vice versa, when $g/\sqrt{N}\gg 1/N$,
the resonance is no more separated from the other modes and a
continuum-like behavior is expected.

\begin{figure}[htbp]
\begin{center}
\includegraphics[height=9cm,angle=-90]{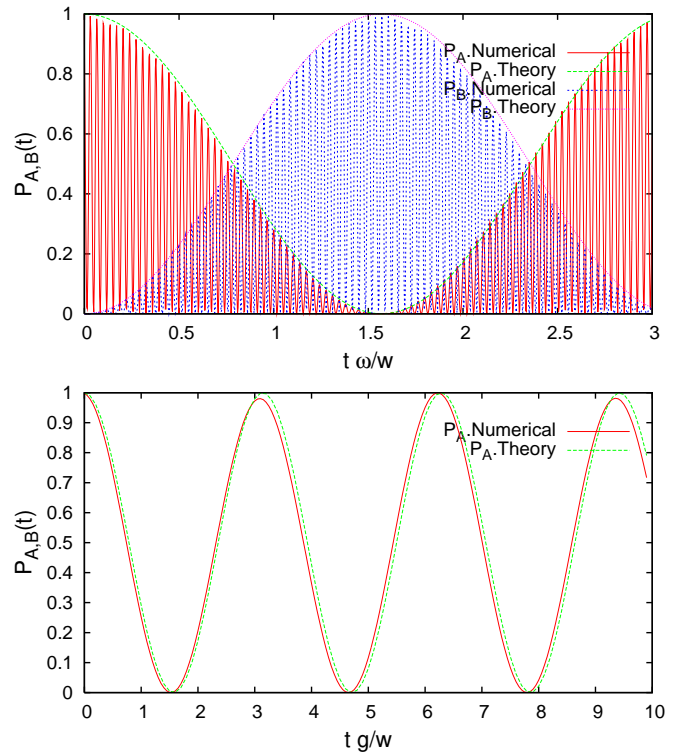}
\end{center}
\caption{ Strong coupling limit: $g=10$, $\Omega=0$, $L=4$, $N=50$.
In the upper panel the low frequency oscillations are compared with
the theory, the time unit is $\omega=g/[2(2g)^L]$. In the lower
panel the same comparison is reported for the higher frequencies.}
\label{fig:fig3}
\end{figure}

The other regime we want to explore is characterized by strong
coupling ($ g\gg 1$). Now, we look for solutions of the order of
$g$,\ then $\Omega $ does not matter, at least in a first
approximation, and can be set to zero. In practice, due to the
strong character of the interaction, the modified energies of
impurities lie ever outside the band, and the channel works in its
continuum limit even for not too large $N$. Considering $\omega \gg
1$, we obtain, by iterative procedure:
\begin{eqnarray}
c_{A}^{\dagger }\left( t\right) =\cos {g}t\left[ \cos
\frac{gt}{2\left( 2g\right) ^{L}}c_{A}^{\dagger }+\left( -1\right)
^{L}i\sin \frac{gt}{2\left( 2g\right) ^{L}}c_{B}^{\dagger }\right]
\nonumber\\+\int dkf\left( k\right) c_{k}^{\dagger },
\end{eqnarray}
where $f\left( k\right) $\ is a function that satisfies the
condition $\int dk\left| f\left( k\right) \right| ^{2}=\sin^2{g}t$.
In this case, we have high frequency oscillations between $A$ and
$B$ and the channel modulated by a low frequency signal which
enables QST. Note that the spectral weight is not entirely
concentrated on the impurities, because at intermediate times the
probability of finding the excitation in the channel is finite. Note
also the different scaling with the distance $L$ of the information
carrying oscillation with respect to weak coupling. In Fig.
\ref{fig:fig3} the probabilities of finding the excitation on $A$
and $B$ are depicted. The lower panel shows the high frequency
oscillation.  The discussion of this limit fails when infinitely
extended discrete spectra are considered, as, for instance, in the
case of finite-length cavities. In this situation, coherent behavior
is expected to come out only from resonance conditions.

In conclusion, we have discussed by analytical and numerical
calculations a number of possible configurations of a quantum bus
allowing perfect state transfer or entanglement generation. The
model we have considered is suitable for implementation of
continuous or discrete quantum channels in different physical
scenarios. We have illustrated in detail in which limits a coherent
behavior emerges, showing that both weak coupling and strong
coupling are suitable for our purposes. We hope that our conclusions
will help to design an experimental setup.

\end{document}